\documentclass[twoside,12pt]{article}  
\usepackage{CJK}   
\usepackage{indentfirst}  
\usepackage{bm}    
\usepackage{graphicx}  

\begin{document}    

\thispagestyle{empty} \vspace*{0.8cm}\hbox
to\textwidth{\vbox{\hfill\huge\sf Chinese Physics B\hfill}}
\par\noindent\rule[3mm]{\textwidth}{0.2pt}\hspace*{-\textwidth}\noindent
\rule[2.5mm]{\textwidth}{0.2pt}


\begin{center}
\LARGE\bf Time evolution of negative binomial optical field in
diffusion channel$^{*}$   
\end{center}

\footnotetext{\hspace*{-.45cm}\footnotesize $^*$Project supported by
the National Basic Research Program of China (Grant No.
2012CB922103), the National Natural Science Foundation of China
(Grant Nos. 11175113, 11274104 and 11404108), and the Natural Science
Foundation of Hubei Province of China (Grant No. 2011CDA021).}
\footnotetext{\hspace*{-.45cm}\footnotesize $^\dag$Corresponding
author. E-mail:  tkliuhs@163.com }

\begin{center}
\rm Liu Tang-Kun$^{\rm a)\dagger}$,\ \ Wu Pan-Pan$^{\rm a)}$, \ \ Shan Chuan-Jia$^{\rm a)}$,
\ \ Liu Ji-Bing$^{\rm a)}$, \ and \ Fan
Hong-Yi$^{\rm b)}$
\end{center}

\begin{center}
\begin{footnotesize} \sl
${}^{\rm a)}$College of Physics and Electronic Science, Hubei Normal University, Huangshi 435002, China\\   
${}^{\rm b)}$Department of Material Science and Engineering,
University of
Science and Technology of China, Hefei 230026, China\\   
\end{footnotesize}
\end{center}

\vspace*{2mm}

\begin{center}
\begin{minipage}{15.5cm}
We find time evolution law of negative binomial optical field in
diffusion channel. We reveal that by adjusting the diffusion
parameter, photon number can controlled.\ Therefore, the diffusion
process can be considered a quantum controlling scheme through
photon addition.
\parindent 20pt\footnotesize
\end{minipage}
\end{center}

\begin{center}
\begin{minipage}{15.5cm}
\begin{minipage}[t]{2.3cm}{\bf Keywords:}\end{minipage}
\begin{minipage}[t]{13.1cm}negative binomial optical field, time evolution, diffusion channel, integration within an
ordered product (IWOP) of operators
\end{minipage}\par\vglue8pt
{\bf PACS: 03.65.-w, 42.50.-p, 02.90.+p}
\end{minipage}
\end{center}

\section{Introduction}  

In a recent paper [1] we have pointed out that an initial number state $%
\left \vert l\right \rangle \left \langle l\right \vert $ undergoing
through
a diffusion channel, described by the master equation [2-3]%
\begin{equation}
\frac{d}{dt}\rho =-\kappa \left( a^{\dagger }a\rho +\rho aa^{\dagger
}-a\rho a^{\dagger }-a^{\dagger }\rho a\right) ,  \label{1}
\end{equation}%
would become a new photon optical field, named
Laguerre-polynomial-weighted
chaotic state, whose density operator is%
\begin{equation}
\rho \equiv \lambda \left( 1-\lambda \right) ^{l}\colon L_{l}\left( \frac{%
-\lambda ^{2}a^{\dag }a}{1-\lambda }\right) e^{-\lambda a^{\dag }a}\colon ,%
\lambda =\frac{1}{1+\kappa t}  \label{2}
\end{equation}%
Here $::$\ denotes normal ordering symbol, $\left \vert l\right
\rangle =a^{\dagger n}\left \vert 0\right \rangle /\sqrt{n!},$
$L_{l}$ is the Laguerre polynomial. Experimentally,\ this new mixed
state may be implemented, i.e., when a number state enters into the
diffusion channel.
Remarkablely, this state is\ characteristic of possessing photon number $%
Tr\left( a^{\dag }a\rho \right) =l+\kappa t$\ at time $t$, so we can
control photon number by adjusting the diffusion parameter $\kappa
,$ this mechanism may provide application in quantum controlling.

\bigskip To go a step further, our aim in this paper is to derive evolution
law of a negative binomial state (NBS) in diffusion channel.
Physically, when an atom absorbs some photons from a thermo light
beam, then the corresponding photon field will be in a negative
binomial state. We are thus challenged by the question: how an
initial NBS evolves in a diffusion channel, what a final state it
will be, and what is the photon number distribution in the final
state. To our knowledge, such questions has not been touched in the
literature before.

Our paper is arranged as follows. In Sec. 2, we convert the density
operator of NBS into normally ordered form. In Sec. 3 based on the
Kraus-operator-solution corresponding to the diffusion channel we
find the evolution law of NBS in diffusion channel. Then in Sec. 4
we calculate the photon number distribution in the final state.

\section{Normally ordered form of the density operator of negative binomial
state}  

Corresponding to the negative-binomial formula
\begin{equation}
\sum_{m=0}^{\infty }\left(
\begin{array}{cc}
m+n\\
m
\end{array}%
\right)
\left( -x\right) ^{m}=\left(
1+x\right) ^{-n-1}  \label{3}
\end{equation}
there exists negative-binomial state of quantum optical field [4]%
\begin{equation}
\rho _{0}=\sum_{n=0}^{\infty }\frac{\left( n+s\right) !}{n!s!}\gamma
^{s+1}\left( 1-\gamma \right) ^{n}\left \vert n\right \rangle \left
\langle n\right \vert ,  0<\gamma<1,  \label{4}
\end{equation}%
where $\left \vert n\right \rangle =a^{\dag n}\left \vert 0\right \rangle /%
\sqrt{n!}$ is the Fock state, $a^{\dag }$ is the photon creation operator, $%
\left \vert 0\right \rangle $ is the vacuum state in Fock space. NBS
is intermediate between a pure thermal state and a pure coherent
state, and its nonclassical properties and algebraic characteristic
have already been studied in detail in Refs. [11, 12]. The photon
number average in this state
is%
\begin{equation}
Tr\left( \rho _{0}a^{\dagger }a\right) =\frac{\left( s+1\right)
\left( 1-\gamma \right) }{\gamma }  \label{5}
\end{equation}%
Using $\left[ a,a^{\dag }\right] =1,$ $a^{s}\left \vert n\right \rangle =%
\sqrt{\frac{n!}{\left( n-s\right) !}}\left \vert n-s\right \rangle
,$ one can reform Eq. (4) as
\begin{eqnarray}
\rho _{0} &=&\frac{\gamma ^{s+1}}{s!\left( 1-\gamma \right) ^{s}}%
a^{s}\sum_{n=0}^{\infty }\left( 1-\gamma \right) ^{n}\left \vert
n\right
\rangle \left \langle n\right \vert a^{\dagger s}  \label{6} \\
&=&\frac{1}{s!n_{c}^{s}}a^{s}\rho _{c}a^{\dagger s}  \nonumber
\end{eqnarray}%
where $\rho _{c}$ denotes a chaotic field
\begin{eqnarray}
\rho _{c} &=&\gamma \sum_{n=0}^{\infty }\left( 1-\gamma \right)
^{n}\left \vert n\right \rangle \left \langle n\right \vert =\gamma
\sum_{n=0}^{\infty }\frac{\left( 1-\gamma \right)
^{n}}{n!}:a^{\dagger n}e^{-a^{\dagger
}a}a^{n}:=\gamma :e^{-\gamma a^{\dagger }a}:  \label{7} \\
&=&\gamma e^{a^{\dagger }a\ln \left( 1-\gamma \right) },  \nonumber
\end{eqnarray}%
Eq. (6) tells us that when some photons are detected for a chaotic
state, e.g. after detecting several photons, the chaotic light field
exhibits
negative-binomial distribution. One can further show $Tr ρ ~_{c}=1$, and%
\begin{equation}
tr\left( \rho _{c}a^{\dagger }a\right) =\frac{1}{\gamma }-1=n_{c}
\label{8}
\end{equation}%
is the mean number of photons of chaotic light field$,$ according to
Bose-Einstein distribution, $n_{c}=\frac{1}{e^{\beta \omega \hbar
}-1},$ here $\beta =\frac{1}{k_{B}T}$, $k_{B}$ is the Boltzmann
constant, $\omega $ is the frequency of chaotic light field. We can
derive the normally ordered form of the density operator of NBS, let
$\ln \left( 1-\gamma \right) =f,$ then $n_{c}=e^{f}/\left(
1-e^{f}\right) $ and by introducing the coherent state
representation $\int \frac{d^{2}z}{\pi }\left \vert z\right \rangle
\left \langle z\right \vert =1,$ $\left \vert z\right \rangle =e^{\frac{%
-\left \vert z\right \vert ^{2}}{2}}e^{za^{\dagger }}\left \vert
0\right \rangle ,$ and employing the\ technique of integration
within an ordered product (IWOP) of operators [5-6] we reform Eq.
(6) as
\begin{eqnarray}
\rho _{0} &=&\frac{1}{s!n_{c}^{s}}a^{s}\rho _{c}a^{\dagger s}=\frac{1-e^{f}}{%
s!n_{c}^{s}}a^{s}e^{fa^{\dagger }a}a^{\dagger s}  \nonumber \\
&=&\frac{1-e^{f}}{s!n_{c}^{s}}\int \frac{d^{2}z}{\pi
}a^{s}e^{fa^{\dagger }a}\left \vert z\right \rangle \left \langle
z\right \vert a^{\dagger s}
\nonumber \\
&=&\frac{1-e^{f}}{s!n_{c}^{s}}\int \frac{d^{2}z}{\pi
}e^{\frac{-\left \vert z\right \vert ^{2}}{2}}a^{s}e^{fa^{\dagger
}a}e^{za^{\dagger }}e^{-fa^{\dagger }a}\left \vert 0\right \rangle
\left \langle z\right \vert
z^{\ast s}  \nonumber \\
&=&\frac{1-e^{f}}{s!n_{c}^{s}}\int \frac{d^{2}z}{\pi
}e^{\frac{-\left \vert z\right \vert ^{2}}{2}}a^{s}e^{za^{\dagger
}e^{f}}\left \vert 0\right
\rangle \left \langle z\right \vert z^{\ast s}  \nonumber \\
&=&\frac{1-e^{f}}{s!n_{c}^{s}}\int \frac{d^{2}z}{\pi }\left(
ze^{f}\right) ^{s}z^{\ast s}\colon e^{-\left \vert z\right \vert
^{2}+za^{\dagger
}e^{f}+z^{\ast }a-a^{\dagger }a}\colon  \nonumber \\
&=&\frac{1-e^{f}}{s!n_{c}^{s}}e^{fs}\colon \sum_{l=0}^{\infty
}e^{^{\left( e^{f}-1\right) a^{\dagger }a}}\frac{\left( n!\right)
^{2}\left( a^{\dagger }ae^{f}\right) ^{n-l}}{l!\left[ \left(
n-l\right) !\right] ^{2}}\colon
\nonumber \\
&=&\left( 1-e^{f}\right) ^{s+1}\colon e^{^{\left( e^{f}-1\right)
a^{\dagger
}a}}L_{s}\left( -a^{\dagger }ae^{f}\right) \colon  \nonumber \\
&=&\gamma ^{s+1}\colon e^{^{-\gamma a^{\dagger }a}}L_{s}\left[
\left( \gamma -1\right) a^{\dagger }a\right] \colon  \label{9}
\end{eqnarray}%
where we have used $\left \vert 0\right \rangle \left \langle
0\right \vert =:e^{-a^{\dagger }a}:,$ and the definition of Laguerre
polynomials
\begin{equation}
L_{s}\left( x\right) =\sum_{l=0}^{s}\frac{\left( -x\right)
^{l}n!}{\left( l!\right) ^{2}\left( n-l\right) !}  \label{10}
\end{equation}%
that (9) is quite different from (1), so they represent different
optical field.

\section{Evolution law of the negative binomial state in diffusion channel}  

Recall in Ref. [7] by using the entangled state representation and
IWOP\
technique we have derived the infinite sum form of $\rho \left( t\right) $%
\begin{eqnarray}
\rho \left( t\right) &=&\sum_{m,n=0}^{\infty
}\sqrt{\frac{1}{m!n!}\frac{\left( \kappa t\right)
^{m+n}}{\left( \kappa t+1\right) ^{m+n+1}}}a^{\dagger m}\left( \frac{1}{%
1+\kappa t}\right) ^{a^{\dagger }a}  \nonumber \\
&&\times a^{n}\rho _{0}a^{\dagger n}\left( \frac{1}{1+\kappa
t}\right)
^{a^{\dagger }a}a^{m}\sqrt{\frac{1}{m!n!}\frac{\left( \kappa t\right) ^{m+n}%
}{\left( \kappa t+1\right) ^{m+n+1}}}  \nonumber \\
&=&\sum_{m,n=0}^{\infty }M_{m,n}\rho _{0}M_{m,n}^{\dagger }
\label{11}
\end{eqnarray}%
where%
\begin{equation}
M_{m,n}=\sqrt{\frac{1}{m!n!}\frac{\left( \kappa t\right)
^{m+n}}{\left( \kappa t+1\right) ^{m+n+1}}}a^{\dagger m}\left(
\frac{1}{1+\kappa t}\right) ^{a^{\dagger }a}a^{n}  \label{12}
\end{equation}%
satisfying $\sum_{m,n=0}^{\infty }M_{m,n}^{\dagger }M_{m,n}=1$,
which is trace conservative.

Now we examine time evolution of negative binomial optical field in
diffusion channel. Substituting $\rho
_{0}=\frac{1}{s!n_{c}^{s}}a^{s}\rho _{c}a^{\dagger s}$ into (11) we
have
\begin{eqnarray}
\rho \left( t\right)  &=&\frac{\gamma ^{s+1}}{s!\left( 1-\gamma \right) ^{s}}%
\sum_{m,n=0}^{\infty }\frac{1}{m!n!}\frac{\left( \kappa t\right) ^{m+n}}{%
\left( \kappa t+1\right) ^{m+n+1}}  \nonumber \\
&&\times a^{\dagger m}\left( \frac{1}{1+\kappa t}\right)
^{a^{\dagger }a}a^{n+s}e^{a^{\dagger }a\ln \left( 1-\gamma \right)
}a^{\dagger s+n}\left( \frac{1}{1+\kappa t}\right) ^{a^{\dagger
}a}a^{m}  \label{13}
\end{eqnarray}%
in which we first consider the summation over $n,$ using%
\begin{equation}
\left( \frac{1}{1+\kappa t}\right) ^{a^{\dagger }a}=e^{-a^{\dagger
}a\ln \left( 1+\kappa t\right) }, e^{fa^{\dagger
}a}ae^{-fa^{\dagger }a}=ae^{-f},  \label{14}
\end{equation}%
we have%
\begin{eqnarray}
&&\sum_{n=0}^{\infty }\frac{1}{n!}\frac{\left( \kappa t\right)
^{n}}{\left( \kappa t+1\right) ^{n}}e^{-a^{\dagger }a\ln \left(
1+\kappa t\right) }a^{n+s}e^{a^{\dagger }a\ln \left( 1-\gamma
\right) }a^{\dagger
s+n}e^{-a^{\dagger }a\ln \left( 1+\kappa t\right) }  \label{15} \\
&=&\left( 1+\kappa t\right) ^{2s}\sum_{n=0}^{\infty }\frac{\left(
\kappa
t\right) ^{n}\left( 1+\kappa t\right) ^{n}}{n!}a^{n+s}e^{a^{\dagger }a\ln %
\left[ \left( 1-\gamma \right) /\left( 1+\kappa t\right) ^{2}\right]
}a^{\dagger s+n}  \nonumber
\end{eqnarray}%
Note from Eq. (9) we have%
\begin{equation}
a^{s}e^{fa^{\dagger }a}a^{\dagger s}=s!e^{fs}\colon e^{\left(
e^{f}-1\right) a^{\dagger }a}L_{s}\left( -a^{\dagger }ae^{f}\right)
\colon   \label{16}
\end{equation}%
it follows%
\begin{eqnarray}
&&a^{n+s}e^{a^{\dagger }a\left[ \ln \left( 1-\gamma \right) -2\ln
\left(
1+\kappa t\right) \right] }a^{\dagger s+n}  \nonumber \\
&=&\left( n+s\right) !e^{\left( n+s\right) \ln \left[ \left(
1-\gamma \right) /\left( 1+\kappa t\right) ^{2}\right] }\colon
e^{\left[ \left( 1-\gamma \right) /\left( 1+\kappa t\right)
^{2}-1\right] a^{\dagger
}a}L_{n+s}\left( -a^{\dagger }a\frac{1-\gamma }{\left( 1+\kappa t\right) ^{2}%
}\right) \colon   \label{17}
\end{eqnarray}%
Substituting (17) into (15) and multiplying $\frac{\gamma
^{s+1}}{\left(
1-\gamma \right) ^{s}s!}$ we see%
\begin{equation}
\begin{array}{c}
\frac{\gamma ^{s+1}\left( 1+\kappa t\right) ^{2s}}{\left( 1-\gamma
\right) ^{s}}\sum_{n=0}^{\infty }\frac{\left( \kappa t+1\right)
^{n}\left( \kappa t\right) ^{n}\left( n+s\right) !}{s!n!}e^{\left(
n+s\right) \ln \left[
\left( 1-\gamma \right) /\left( 1+\kappa t\right) ^{2}\right] }\colon e^{%
\left[ \left( 1-\gamma \right) /\left( 1+\kappa t\right)
^{2}-1\right]
a^{\dagger }a}L_{n+s}\left( -a^{\dagger }a\frac{\left( 1-\gamma \right) }{%
\left( 1+\kappa t\right) ^{2}}\right) : \\
=\gamma ^{s+1}\sum_{n=0}^{\infty }\frac{\left( n+s\right) !\left(
\kappa
t\right) ^{n}\left( 1-\gamma \right) ^{n}}{s!n!\left( \kappa t+1\right) ^{n}}%
\colon L_{n+s}\left( -a^{\dagger }a\frac{\left( 1-\gamma \right)
}{\left( 1+\kappa t\right) ^{2}}\right) e^{\left[ \left( 1-\gamma
\right) /\left(
1+\kappa t\right) ^{2}-1\right] a^{\dagger }a}:%
\end{array}
\label{18}
\end{equation}%
Then we use the new generating function formula about the Laguerre
polynomials [8]$.$
\begin{equation}
\sum_{n=0}^{\infty }\frac{\left( n+s\right) !\left( -\lambda \right) ^{n}}{%
n!s!}L_{n+s}\left( z\right) =\left( 1+\lambda \right) ^{-s-1}e^{\frac{%
\lambda z}{1+\lambda }}L_{s}\left( \frac{z}{1+\lambda }\right) .
\label{19}
\end{equation}%
we obtain%
\begin{eqnarray}
(18) &=&\left[ \frac{\gamma \left( \kappa t+1\right) }{1+\kappa t\gamma }%
\right] ^{s+1}:L_{s}\left( -a^{\dagger }a\frac{\left( 1-\gamma \right) }{%
\left( 1+\kappa t\gamma \right) \left( 1+\kappa t\right) }\right)
e^{a^{\dagger }a\frac{\kappa t\left( 1-\gamma \right) ^{2}}{\left(
1+\kappa t\gamma \right) \left( 1+\kappa t\right) ^{2}}}e^{\left[
\left( 1-\gamma \right) /\left( 1+\kappa t\right) ^{2}-1\right]
a^{\dagger }a}:  \label{20}
\\
&=&\left[ \frac{\gamma \left( \kappa t+1\right) }{1+\kappa t\gamma
}\right] ^{s+1}:L_{s}\left( -a^{\dagger }a\frac{\left( 1-\gamma
\right) }{\left(
1+\kappa t\gamma \right) \left( 1+\kappa t\right) }\right) e^{\left[ -\frac{%
\gamma -1}{\left( t\kappa +1\right) \left( t\kappa \gamma +1\right) }-1%
\right] a^{\dagger }a}:  \nonumber
\end{eqnarray}%
For $\rho \left( t\right) $ in Eq. (13) It remains to perform
summation over $m,$ using the summation technique within normal
ordering we have
\begin{eqnarray}
\rho \left( t\right)  &=&\left[ \frac{\gamma \left( \kappa t+1\right) }{%
1+\kappa t\gamma }\right] ^{s+1}\sum_{m=0}^{\infty }\frac{\left(
\kappa
t\right) ^{m}}{m!\left( \kappa t+1\right) ^{m+1}}  \label{21} \\
\times  &:&a^{\dagger m}e^{\left[ -\frac{\gamma -1}{\left( t\kappa
+1\right) \left( t\kappa \gamma +1\right) }-1\right] a^{\dagger
}a}L_{s}\left( -a^{\dagger }a\frac{\left( 1-\gamma \right) }{\left(
1+\kappa t\gamma
\right) \left( 1+\kappa t\right) }\right) a^{m}:  \nonumber \\
&=&C:e^{Ea^{\dagger }a}L_{s}\left( a^{\dagger }aF\right) :
\nonumber
\end{eqnarray}%
where%
\begin{equation}
E=-\frac{\gamma }{\kappa t\gamma +1}, F=\frac{\gamma
-1}{\left(
1+\kappa t\gamma \right) \left( 1+\kappa t\right) },F+E=\frac{-1}{%
1+\kappa t}  \label{22}
\end{equation}%
\begin{equation}
C=\frac{1}{1+\kappa t}\left[ \frac{\gamma \left( \kappa t+1\right) }{%
1+\kappa t\gamma }\right] ^{s+1}  \label{23}
\end{equation}%
Comparing $\rho \left( t\right) $ in (21) with $\rho _{0}$ in (9) we
can see the big difference. Now we must check if $Tr\rho \left(
t\right) =1,$ in \
fact, using the coherent state's completeness relation $1=\int \frac{d^{2}z}{%
\pi }\left \vert z\right \rangle \left \langle z\right \vert $ and
\begin{equation}
\int_{0}^{\infty }e^{-bx}L_{l}\left( x\right) dx=\left( b-1\right)
^{l}b^{-l-1}.  \label{24}
\end{equation}%
we do have%
\begin{eqnarray}
Tr\rho \left( t\right)  &=&CTr\left[ :e^{Ea^{\dagger }a}L_{s}\left(
a^{\dagger }aF\right) :\int \frac{d^{2}z}{\pi }\left \vert z\right
\rangle
\left \langle z\right \vert \right]   \nonumber \\
&=&C\int \frac{d^{2}z}{\pi }e^{-\frac{\gamma }{t\kappa \gamma +1}%
|z|^{2}}L_{s}\left( -\frac{|z|^{2}\left( 1-\gamma \right) }{\left(
1+\kappa
t\gamma \right) \left( 1+\kappa t\right) }\right)   \nonumber \\
&=&C\frac{\left( 1+\kappa t\gamma \right) \left( 1+\kappa t\right)
}{\left( \gamma -1\right) }\int_{0}^{\infty }dr^{\prime
}e^{-\frac{\gamma \left( 1+\kappa t\right) r^{\prime }}{\left(
\gamma -1\right) }}L_{s}\left(
r^{\prime }\right)   \label{25} \\
&=&C\frac{\left( 1+\kappa t\gamma \right) \left( 1+\kappa t\right)
}{\gamma -1}\left( \frac{\gamma \kappa t+1}{\gamma -1}\right)
^{s}\left[ \frac{\gamma -1}{\gamma \left( 1+\kappa t\right) }\right]
^{s+1}=1  \nonumber
\end{eqnarray}

\section{Photon number average in the final state}

Now we evalute photon number average in the final state, using (21),
(10)
and (22)-(24) we have%
\begin{eqnarray}
Tr\left[ \rho \left( t\right) a^{\dagger }a\right] &=&C\int \frac{d^{2}z}{%
\pi }\left \langle z\right \vert a^{\dagger }a:e^{Ea^{\dagger
}a}L_{s}\left(
a^{\dagger }aF\right) :\left \vert z\right \rangle  \label{26} \\
&=&C\int \frac{d^{2}z}{\pi }z^{\ast }\left \langle z\right \vert \left \{ %
\left[ a,:e^{Ea^{\dagger }a}L_{s}\left( a^{\dagger }aF\right)
:\right] +:e^{Ea^{\dagger }a}L_{s}\left( a^{\dagger }aF\right)
:a\right \} \left
\vert z\right \rangle  \nonumber \\
&=&C\int \frac{d^{2}z}{\pi }\left \langle z\right \vert z^{\ast
}\left \{ \frac{\partial }{\partial a^{\dagger }}:e^{Ea^{\dagger
}a}L_{s}\left( a^{\dagger }aF\right)
:+|z|^{2}e^{E|z|^{2}}L_{s}\left( |z|^{2}F\right)
\right \} \left \vert z\right \rangle  \nonumber \\
&=&C\int \frac{d^{2}z}{\pi
}e^{E|z|^{2}}\sum_{l=0}^{s}\frac{l|z|^{2l}\left(
-F\right) ^{l}s!}{\left( l!\right) ^{2}\left( s-l\right) !}+C\int \frac{%
d^{2}z}{\pi }\left( 1+E\right) |z|^{2}e^{E|z|^{2}}L_{s}\left(
|z|^{2}F\right)
\nonumber \\
&=&C\sum_{l=0}^{s}\frac{\left( -F\right) ^{l}s!}{\left( l!\right)
^{2}\left( s-l\right) !}\left[ l\int \frac{d^{2}z}{\pi
}e^{E|z|^{2}}|z|^{2l}+\left(
1+E\right) \int \frac{d^{2}z}{\pi }|z|^{2\left( l+1\right) }e^{E|z|^{2}}%
\right]  \nonumber \\
&=&C\sum_{l=0}^{s}\frac{\left( -F\right) ^{l}s!}{\left( l!\right)
^{2}\left( s-l\right) !}\left[ l\frac{l!}{\left( -E\right)
^{l+1}}+\left( 1+E\right)
\frac{\left( l+1\right) !}{\left( -E\right) ^{l+2}}\right]  \nonumber \\
&=&tk+\frac{\left( s+1\right) \left( 1-\gamma \right) }{\gamma }
\nonumber
\end{eqnarray}%
Comparing with Eq. (5) we see that after passing through a diffusion
channel, the photon average of a NBS
varies from $\frac{\left( s+1\right)
\left( 1-\gamma \right) }{\gamma }$ to $tk+\frac{\left( s+1\right)
\left( 1-\gamma
\right) }{\gamma },$%
\begin{equation}
Tr\left[ \rho (t)a^{\dagger }a\right] =tk+\frac{\left( s+1\right)
\left( 1-\gamma \right) }{\gamma }=tk+Tr\left( \rho _{0}a^{\dagger
}a\right) \label{27}
\end{equation}\\

This result is encourageous, since by adjusting the
diffusion parameter $\kappa $, we can control photon number, when
$\kappa $\ is small,\ it slightly increases by an amount $\kappa
t.$\ Therefore, this diffusion process for NBS can be considered a quantum
controlling scheme through photon addition.


\begin{thebibliography}{99}
\bibitem{1} Fan HongYi, Lou SenYue and Pan XiaoYin 2014
SCIENCE CHINA-PHYSICS MECHANICS \& ASTRONOMY \textbf{57} 1649-1653
\bibitem{2} Carmichael H J 1999 Statistical Methods in Ouantum Optics I, Master
Equation and Fokker-Planck equations (Berlin: Springer-Verlag)
\bibitem{3} Orszag M 2000 Quantum Optics (Berlin:Springer-Verlag)
\bibitem{4} G. S. Agarwal 1992 Phys. Rev. A \textbf{45} 1787
\bibitem{5} Fan H Y, H. L. Lu and Y. Fan 2006 Ann. Phys. , \textbf{321} 480-494
\bibitem{6} Fan H Y 2003 J. Opt. B: Quantum Semiclass. Opt. , \textbf{5} R147-R163
\bibitem{7} Liu Tang-Kun, Shan Chuan-Jia, Liu Ji-Bing, and Fan Hong-Yi 2014 Chin.
Phys. B \textbf{23} 030303
\bibitem{8} Fan Hong-yi, Lou Sen-yue, Pan Xiao-yin and Da Chen 2013 Acta Phys.
Sin. \textbf{62}  240301(in Chinese)


\end{thebibliography}
\end{document}